# Unsupervised Machine Learning and EMF radiation in schools: a study of 205 schools in Greece.

**Yiannis Kiouvrekis[1,2]  (kiouvrekis.y@uth.gr, yiannis.kiouvrekis@gmail.com ) , Aris Alexias[1], Yiannis Filipopoulos[2,3], Vasiliki Softa[1], Ch. D. Tyrakis[1], C. Kappas[1]**

[1]*Department of Medical Physics, University of Thessaly, Larissa, Hellas*

[2]*University of Nicosia, Nicosia, Cyprus*

[3]*Hellenic American University, New Hampshire USA*

**Address for Correspondence:**

**Yiannis Kiouvrekis**

**University of Thessaly, Larissa, Hellas**

**Phone: +30 6944347515**

**E- mail:** kiouvrekis.y@uth.gr – yiannis.kiouvrekis@gmail.com

**Abstract.** The expansion of network infrastructure in Greece has raised concerns about the possible negative health effects on sensitive groups, such as children, from exposure to long-term radiofrequency electromagnetic fields (RF-EMFs). The objective of this study is to apply Unsupervised Machine Learning methods such as hierarchical clustering, in order to establish patterns of EMF radiation in schools. To this end, we performed measurements in the majority schools units (n=205) in the region of Thessaly in order to calculate the mean value for RF-EMF exposure in the 27 MHz–3 GHz range, which covers the whole spectrum of RF-EMF sources. Hierarchical clustering dendrogram analysis shows that population density in urban areas of Thessaly bears no relation to the level of EMF exposure in schools. Furthermore, in 97.5% of schools found in the Thessaly region, the exposure level is at least 3500 times below the Greek exposure limits while in 2.5% it is a little less than 500 times below the limit.

**Introduction**

The increasing use of devices powered by electromagnetic radiation technology has led to a respective increase of public concern over potential health hazards. These concerns involve the effects of electromagnetic radiation on public health in general and on the health of sensitive groups in particular.

This is a multifactor issue; on the one hand, studies are conducted on the effects of non-ionizing radiation on health and, on the other hand, there is an ongoing research on electromagnetic radiation levels in different environments.

A brief literature review highlights the multifactor nature of the issue [1]. Studies involving electromagnetic radiation measurements can be conducted either in indoor environments [2,3,4,5] or outdoor environments [6,7], they are typically conducted in public environments, i.e. workplaces [8,9], educational facilities [10, 11, 12, 13, 14], public transport [15, 16], etc.

There are several assessment strategies. These include (a) on site (in situ) measurements of EMF levels in locations selected with respect to broadband availability, (b) measurements concerning specific locations to record an average value – values are recorded during a certain time span – with respect to the sampling time used by the measuring device [27], (c) personal exposimetry [18, 19] and (d) measurements conducted through use of modelling and simulation [27].

In previous research studies [23, 24, 25, 26], two separate sampling methodologies were followed in order to establish an approximation of the EMF radiation value in Greek schools. This study's innovation is to use a geographic information system (GIS) with a view to a different approach in collecting, storing, analysing and presenting information with respect to assessing electromagnetic radiation in schools. In this manner we will be able to present the data on a geographic map and provide graphic depictions of relations between spatial data.

The necessity of this type of analysis arises out of the need to identify areas with an increased electromagnetic radiation value. This is poorly understood so far, in part due to a) the heterogeneous geographic structure of Greece, which often poses difficulties in distinguishing between urban and semi-urban areas and b) the great variety in population density distribution, as a feature of urban areas.

**Methodology – Analysis**

In order to identify the spot with the highest power density, a broadband survey was conducted. In accordance with the Greek Atomic Energy Commission's guidelines, Greek legislation as well as international organizations' standards we followed the following procedure: The instrument we used is an SRM 3006 spectrum analyzer (Narda Safety Tests Solutions, Pfullingen, Germany) with a detection range of 27 MHz - 3 GHz. By applying the function "Safety Evaluation" of the instrument, the assessment of the electric field strength was performed by measuring the electric field intensity (in V/m) over the entire frequency range of 27MHz to 3GHz and by obtaining frequency selective measurements over a time interval of 6min at three successive heights that correspond to an exposed human body: 110cm, 150cm and 170cm above the ground. In those selected spots, the user measures and records the average EMF strength. In order to assess uncertainty, all factors which can introduce errors in field measurements, whether internal or external to the measurement method, were taken into account.

The data set contains 6 variables: *Area, Exists, Electric Field Strength, Dense; Ratio;* and *Limit*. *Area* is a categorical variable with 4 groups (cities of Larisa, Karditsa, Trikala and Volos); *Exists* is a categorical variable with 2 groups (YES = If the school is fewer than 300 metres from the closest antenna NO = if the school is more than 300 metres away from the closest antenna); *Electric Field Strength* is a continuous variable in V/m; *Dense* is a continuous variable (population density) persons/km^2. *Ratio* is a continuous variable (the level of exposure ratio) and *Limit* is another continuous variable (how much below or above the limit is the value of the Electric Field Strength.

For the purposes of our analysis we will utilize Clustering algorithms. These algorithms employ a technique which *groups similar data points together* so that points belonging to the **same group** are more **similar** to each other compared to points belonging to other groups. Each group containing similar points is called a **Cluster**. To be more precise, we will use the Hierarchical Clustering Technique algorithm, which follows one of two methodologies:

1. Methodology 1: Agglomerative

    Each point is considered an isolated cluster. In every loop, similar clusters are merged together with other clusters in order to form one cluster.

    **Algorithm development:**

    STEP 1: **Assess** proximity of isolated points and consider them as isolated clusters / **Allows** for every point to be a cluster.

STEP 2: **Merge together** similar clusters and **form** one cluster.

STEP 3: **Re-calculate** proximity of new clusters / **Merge together** similar clusters.

STEP 4: **Merge together** all clusters and **form** a single cluster.

This technique can be depicted by employing a Tree Graph (a graph resembling the braches of a tree, which records sequences of merges and breaking ups.

2. Methodology 2: Divisive
   This technique is the reverse of technique 1. We consider all data points as a single cluster and during each loop we single out those points which are not similar to other points inside the cluster. Thus, every point we single out is considered a isolated cluster until we finally end up with $n$ clusters, which are equal to the data points in number.

How do we assess similarity between two clusters? This is an important step, since agglomeration of division of groups is dependent upon it. Therefore, in order to assess similarity between clusters, we use the following approaches:

- *Minimum or Single clustering*
    - Can divide non-elliptic shapes, as long as the distance between clusters is not short.
    - Cannot divide clusters correctly, if there is noise between clusters.

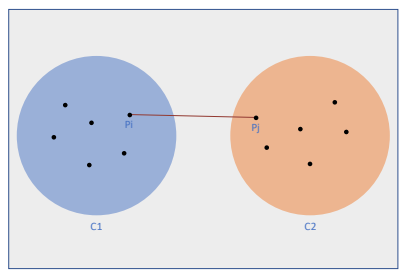

- *Maximum or Complete clustering*
    - Can divide clusters correctly, if there is noise between clusters.
    - Does not work well on spherical clusters.
    - Tends to break up big clusters.

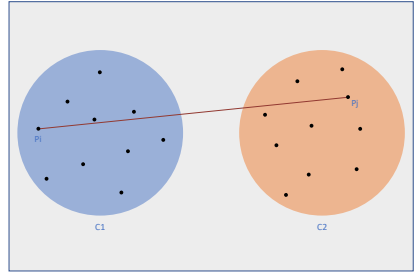

- *Average or Mean clustering*
    - Takes into account all paired points and assesses their similarities and their similarity average.
    - Can divide clusters correctly, if there is noise between clusters.
    - Does not work well on round clusters.

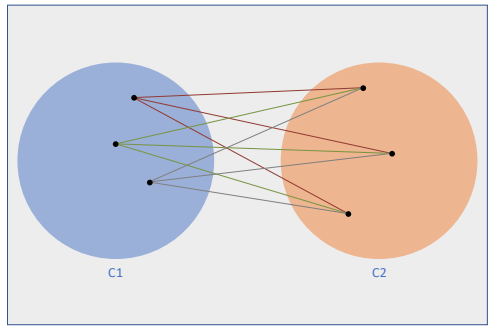

- *Distance between centroids*
    - Establishes the centroids of two clusters and establishes the similarity between the two centroids as the similarity between the two clusters.

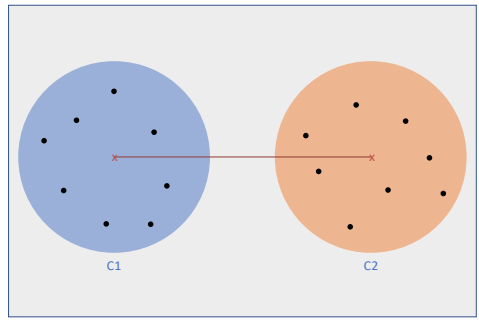

- *Ward's method*
    - This method follows the same procedure as the method of calculating the similarity of a cluster average. Furthermore, calculates the squared sum of the interval between two points.
    - Can divide clusters correctly, if there is noise between clusters.
    - Does not work well on spherical clusters.

Since all measurements of our data set are available, this algorithm, when applied, will initially treat measurements as isolated and will subsequently group them according to their similarities.

This enables us to create an aggregate "virtual" map to compare all data and thus draw conclusions concerning measurements, e.g. which of them are nearing the limit and which are lower or significantly lower than the limit.

**Results**

Data from 205 urban area schools in the Region of Thessaly were uploaded to QGIS resulting in the following map:

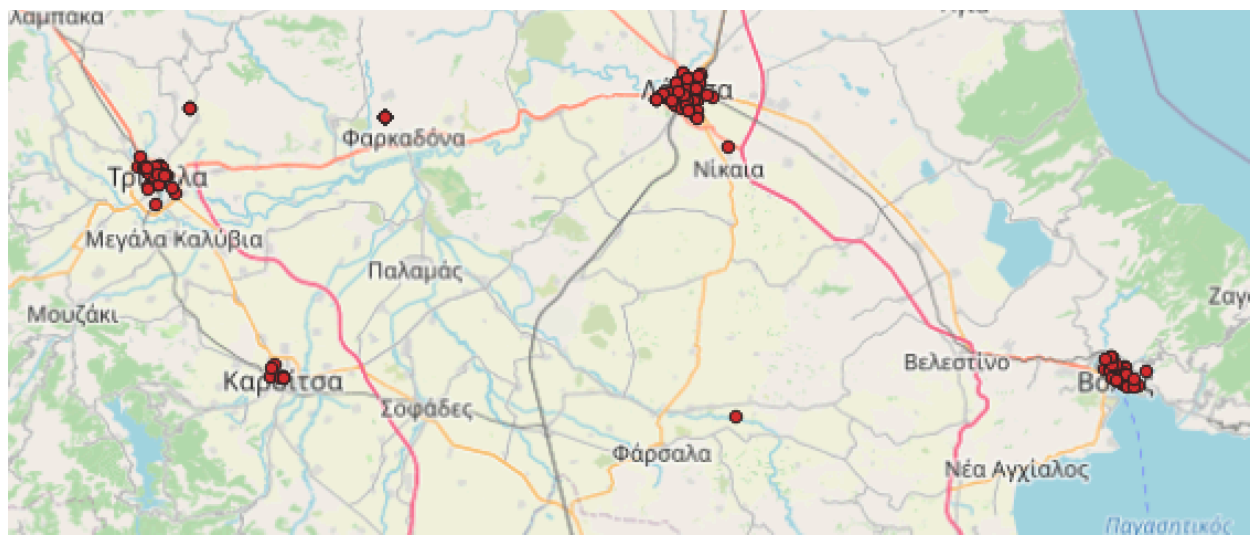

Figure 1 Χαρτης Σχολείων

In Figures 2a and 2b we can see the distribution of schools per city as well as the distribution of schools in relation to whether the nearest antenna is found within 300m or not.

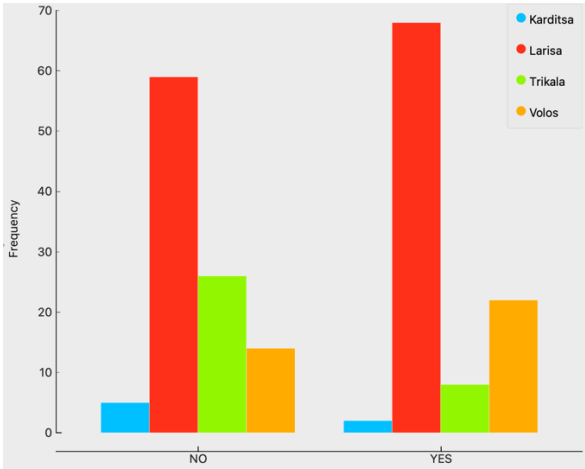 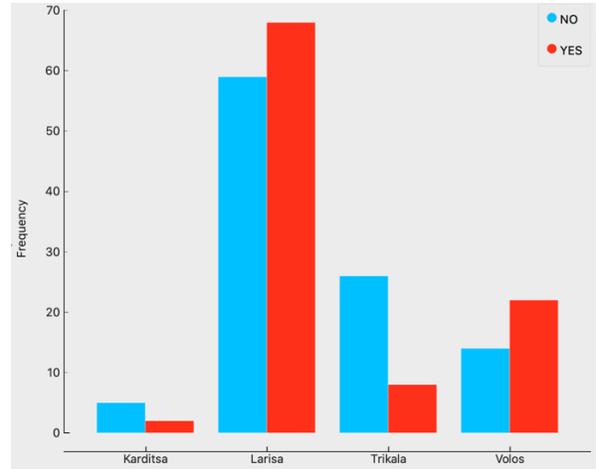

Figure 2 a                    Figure 3 b

After calculating the standard measure values, we come up with the following: the minimum value of Electric Field Strength is 0.2731 V/n, 1st (lower) quartile (Q1) is 0.2962 V/m, median M = 0.3035 , 3rd (upper) quartile (Q3) is 0.3710 V/m, and the maximum value is 1.9210V/m;. The mean is equal to 0.3565 and the median M = 0.3035 are also different; this indicates that the data are skewed as we can verify from the histogram.

**Hierarchical Clustering**

At first, we standardized the variables so as to render them comparable pairs; afterwards, we performed agglomerative hierarchical clustering. The first step was to calculate the agglomerative coefficients (table 1), which measure the amount of clustering structure found while the second step was to compute the dissimilarity matrix using the Euclidean metric.

|  | Complete method | Average method | Single method | Ward method |
|---|---|---|---|---|
| agglomerative coefficient | 0.9893483 | 0.9880241 | 0.9906571 | 0.9920184 |

Since we understood in advance that in the interests of the clustering process we had to select the optimal number of clusters, we followed the average silhouette method; the following table shows that the optimal number of clusters is 5.

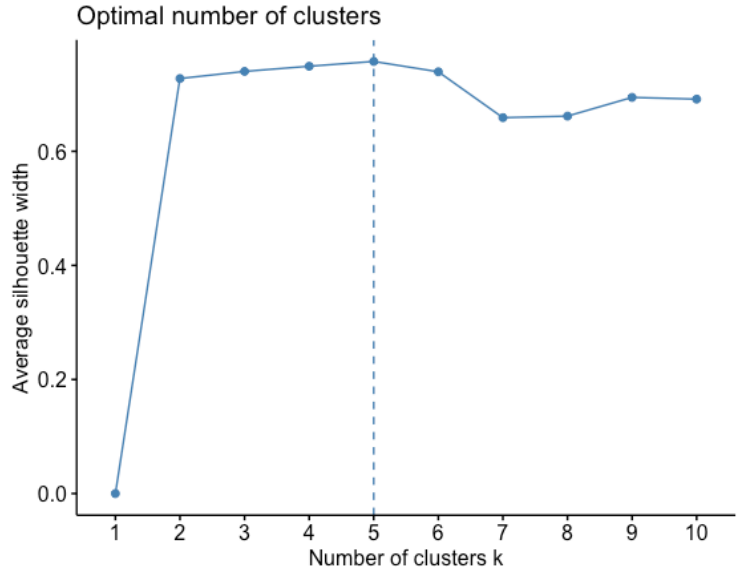

Figure 3

A clustering following different methods is presented in the following Figure.

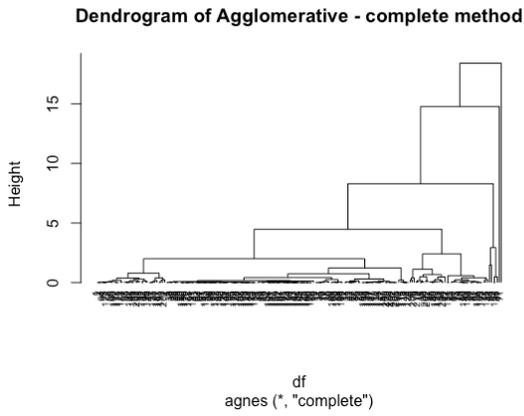 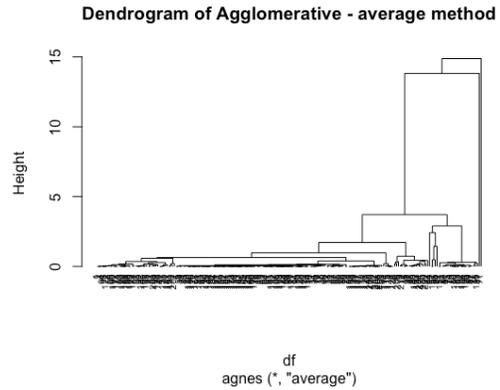

Figure 4 a                                   Figure 4 b

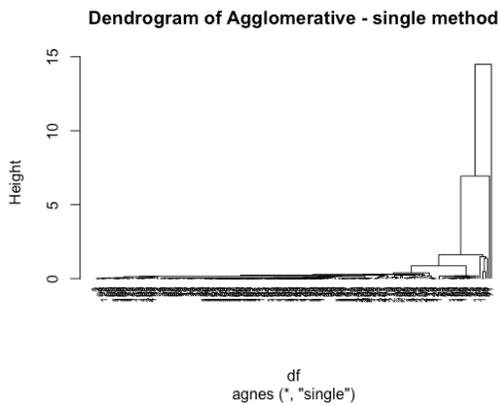 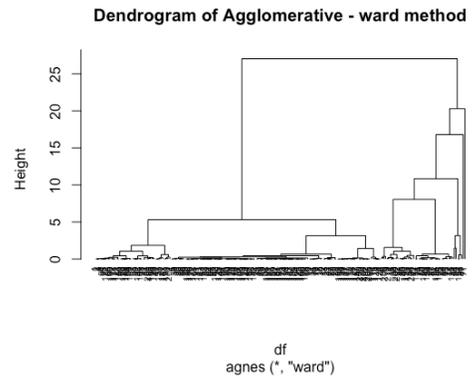

Figure 4 c                                              Figure 4 d

In addition, we can compare two dendrograms. The outputs (images) show "unique" nodes exhibiting a combination of labels/objects not found in the other dendrogram, represented by dotted lines. The alignment quality between the two dendrograms can be measured with the entanglement function. Entanglement is a value between 1 and 0. A lower entanglement coefficient corresponds to a good alignment.

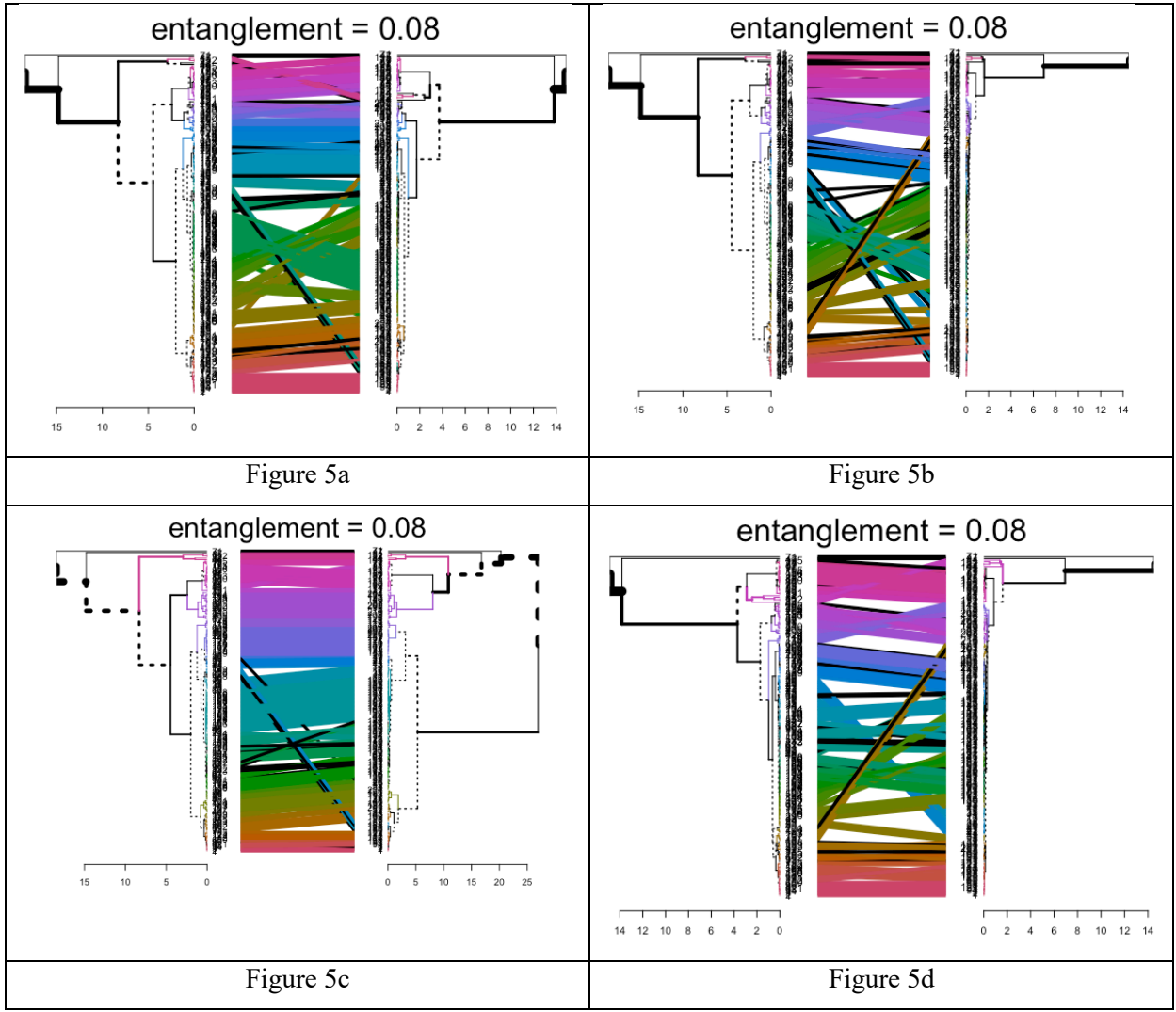

| Figure 5a | Figure 5b |
|---|---|
| Figure 5c | Figure 5d |

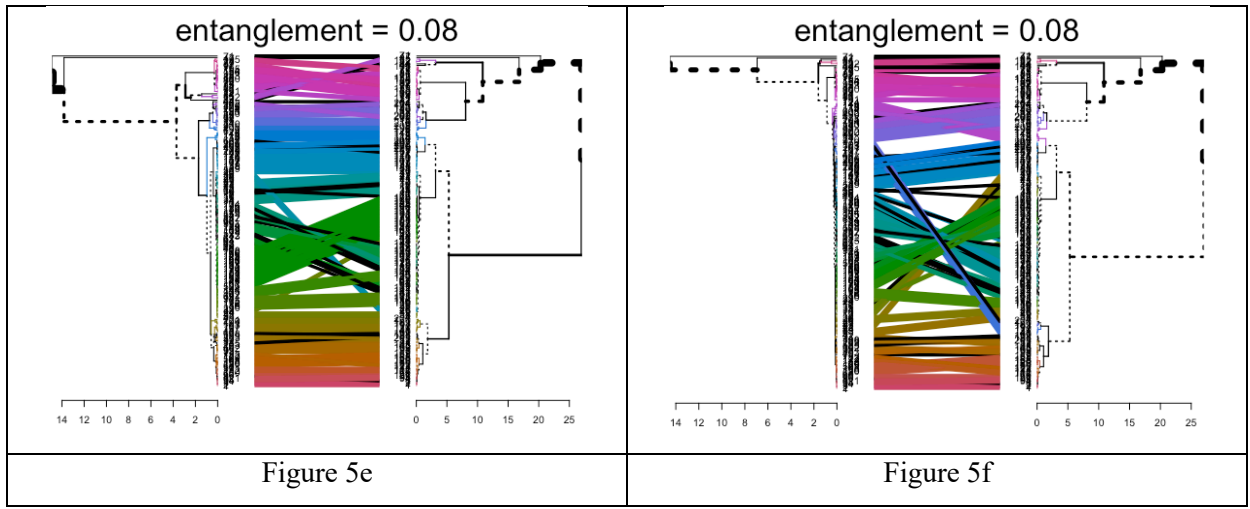

| Figure 5e | Figure 5f |

## Discussion

The Figures presented above were produced using the R programming language. After a thorough analysis and an application of hierarchical clustering, we conclude that there is no specific pattern in total RMF strength in the 27MHz - 3GHz range in schools situated in urban areas of the region of Thessaly. Furthermore, the mean value recorded concerning the 205 schools included in our research is lower than the values recorded in [24, 25]. Hierarchical clustering dendrogram analysis shows that population density in urban areas of Thessaly bears no relation to the level of EMF exposure in schools. According to our measurements in 97.5% of schools found in the Thessaly region, the exposure level is at least 3500 times below the Greek exposure limits while in 2.5% it is a little less than 500 times below the limit.

This is a very significant finding, as it indicates that a high EMF strength in the 27 MHz–3 GHz range is not recorded in schools of any area category as large as that of the region of Thessaly in Greece.

We should note that, in Greece the RF limit values for sensitive land uses (schools, hospitals etc.) are set at 60% of those recommended by the EU standard and 70% for the general population.